\documentclass[twocolumn,english,twocolumn,prb,amssymb,amsmath,showpacs,superscriptaddress]{revtex4-1}

\usepackage[latin9]{inputenc}
\usepackage{graphicx,color}
\usepackage{amssymb}
\usepackage{amsmath}
\usepackage[bookmarks]{hyperref}
\usepackage{babel}
\usepackage{verbatim}
\usepackage{selinput}
\usepackage{ulem}

\newcommand{\be}{\begin{equation}}
\newcommand{\ee}{\end{equation}}

\newcommand{\ket}[1]{\left|#1\right\rangle}

\begin{document}

\title{Stretched exponential decay of Majorana edge modes in many-body localized
Kitaev chains under dissipation}
\author{Alexander Carmele$^{\S}$}
\email{alex@itp.tu-berlin.de}
\affiliation{Institute for Quantum Optics and Quantum Information of the
Austrian Academy of Sciences, A-6020 Innsbruck, Austria}
\affiliation{Institut f\"ur Theoretische Physik, Technische Universit\"at
Berlin, Hardenbergstra\ss e 36, 10623 Berlin, Germany}
\author{Markus Heyl$^{\S}$} 
\email{markus.heyl@uibk.ac.at}
\thanks{\\$^{\S}$ These authors contributed equally to this work.}
\affiliation{Institute for Quantum Optics and Quantum Information of the
Austrian Academy of Sciences, A-6020 Innsbruck, Austria}
\affiliation{Institute for Theoretical Physics, University of Innsbruck, A-6020
Innsbruck, Austria}
\affiliation{Physik Department, Technische Universit\"at M\"unchen, 85747 Garching, Germany}
\author{Christina Kraus}
\affiliation{Institute for Quantum Optics and Quantum Information of the
Austrian Academy of Sciences, A-6020 Innsbruck, Austria}
\affiliation{Institute for Theoretical Physics, University of Innsbruck, A-6020
Innsbruck, Austria}
\author{Marcello Dalmonte}
\email{marcello.dalmonte@uibk.ac.at}
\affiliation{Institute for Quantum Optics and Quantum Information of the
Austrian Academy of Sciences, A-6020 Innsbruck, Austria}
\affiliation{Institute for Theoretical Physics, University of Innsbruck, A-6020
Innsbruck, Austria}
\begin{abstract}

We investigate the resilience of symmetry-protected topological edge states at the boundaries of 
Kitaev chains
 in the presence of a bath which explicitly
introduces symmetry-breaking terms. Specifically, we focus on single-particle losses and gains, violating the protecting parity
symmetry, which could generically occur in realistic scenarios. For homogeneous
systems, we show that the Majorana mode decays exponentially fast. 
By the inclusion of strong disorder, where the closed system enters a many-body localized phase, we find that the Majorana mode can be stabilized substantially. The decay of the Majorana converts into a stretched
exponential form for particle losses or gains occuring in the bulk. In particular, for pure loss dynamics we find a universal exponent $\alpha \simeq 2/3$. We show that this holds both in the
Anderson and many-body localized regimes. Our results thus provide a first 
step to stabilize edge states even in the presence of 
symmetry-breaking
environments.

\end{abstract}


\maketitle


\section{Introduction} 

Symmetry-protected topological states are a well established class of phases
of matter which encompasses a rich variety of microscopic incarnations, ranging
from spin chains to fermionic 
systems~\cite{wenbook,PhysRevB.83.075102,PhysRevB.81.134509}. 
These phases of matter are a direct manifestation of a
(non-local) symmetry in a model Hamiltonian, first identified in the context of
the Haldane phase of spin-1 chains~\cite{Kennedy:1992kq}, which can give rise to
topological
order and zero energy modes at the sample boundaries.  A prominent example of
this mechanism is provided by the Kitaev chain~\cite{kitaev_chain} which exhibits Majorana
edge states~\cite{Beenakker:2011yg} once defined on
a open geometry. These edge modes, which have anyonic statistics and can be
potentially useful for quantum information purposes~\cite{Nayak:2008qr}, have a
direct relation with
a global $\mathbb{Z}_2$ parity symmetry of certain one-dimensional
superconductors~\cite{kitaev_chain}, usually granted by the proximity effect.
However, naturally occurring dissipation can explicitly break the protecting
microscopic
symmetry~\cite{mazza_robustness_majorana,bardyn_topology_dissipation}. In this
context the major challenge is to stabilize the Majorana modes by exploiting 
additional mechanisms which could counteract the drastic action of symmetry
breaking terms in open quantum 
system settings.

In this work, we investigate the resilience of Majorana edge modes in the
presence of incoherent symmetry-breaking dissipation. We show that strong
disorder can lead to an exponential gain in the edge mode's stability compared
to the case of a homogeneous system. While in a homogeneous system we show that
the temporal decay is exponential, in presence of disorder we find that the Majorana mode occupation decays in a
{\it stretched} exponential form. Remarkably, when the coupling to the
environment induces pure particle losses, the associated exponent $\alpha \simeq
2/3$ is universal and independent of any microscopic details within our numerical
accuracy. 

The effect of disorder on the time-dependent dynamics of closed quantum
many-body systems has been actively investigated in both non-interacting and
interacting scenarios. For noninteracting particles in low dimensions,
disorder leads to Anderson localization (AL): all single-particle eigenstates
become localized immediately for infinitesimal
disorder~\cite{Anderson1958,Abrahams1979}. Recently, the concept of AL has been
generalized to interacting systems in the context of many-body localization
(MBL)~\cite{Altshuler1997hx,basko_stabilization_disorder_interacting,Nandkishore2014,Altman2014}. Importantly, compared to
quantum many-body systems in thermodynamic phases, MBL systems display
unconventional properties~\cite{Znidaric2008,Bardarson2012,Altman2014,Nandkishore2014}. In particular, this
encompasses the possibility of phase transitions at nonzero temperature in one
dimension~\cite{basko_stabilization_disorder_interacting,Aleiner2010,
Oganesyan2007nm,Pal2010}. Moreover, it has been shown that topological
properties, in one-dimensional thermodynamic systems only 
present in ground states, can survive over all of the
spectrum~\cite{Huse2013,Bauer2013,Chandran2014}, greatly enhancing the regime of
stability for topological order in closed systems~\cite{Bahri2013}. 

Motivated by these findings, we address in this work the fundamental question if
and how
the MBL mechanism can affect the stability of topological order, and in
particular the resilience of edge modes, when the system is coupled to an
incoherent bath.
This question finds natural applications in diverse systems, such as solid state
devices and cold atom settings, which have inherent sources of dissipation.
Our results support the conclusion that MBL qualitatively enhances the
resilience of edge states in symmetry-protected topological states even in the
presence of the worst-case scenario of a bath inducing symmetry-breaking
perturbations.

\begin{figure}
\centering
\includegraphics[width=0.7\linewidth]{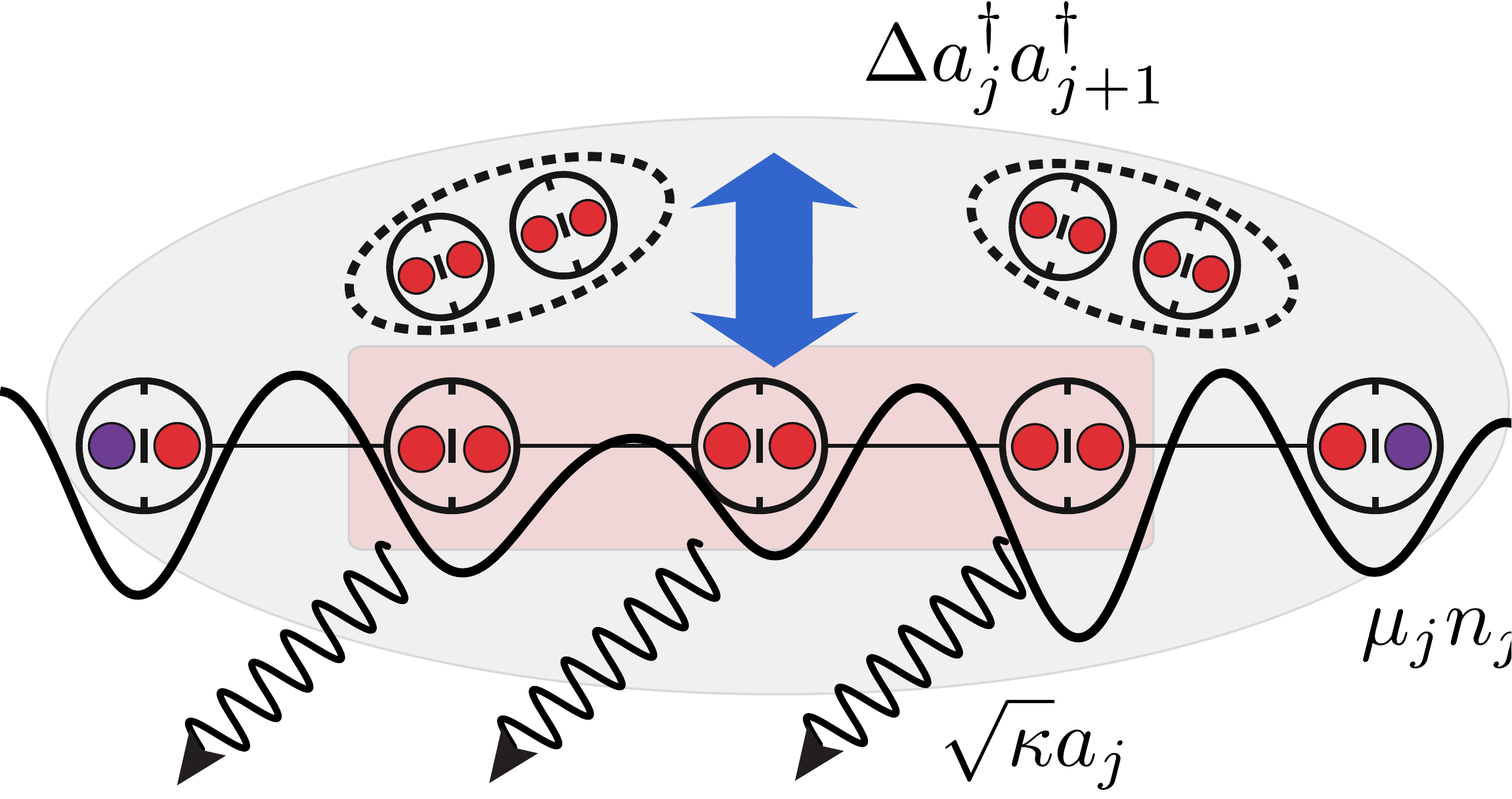}
\caption{(color online) Schematic picture of the dynamics of a disordered Kitaev
chain subject to a bath. The system is initialized in the ground state of the
ideal Kitaev chain, where it is better visualized splitting the fermionic degree
of freedom at each site into a pair of Majorana fermions (small full circles).
In the ground state, all Majorana fermions in the middle couple over bonds,
leaving the first and last Majorana fermion uncoupled (blue circled). The
dynamics is the subject to a disordered potential (black, thick line) in
addition to pairing to a superconductor (grey shaded area) and to the action of
a bath acting on the bulk (red area). The presence of the bath explicitly breaks
the parity symmetry of the wire due to particle loss. }
\label{fig:1}
\end{figure}

In concrete, we study the combined effects of disorder and
interactions in a Kitaev chain coupled to a Markovian bath which explicitly
breaks the parity symmetry of the system. The setup we have in mind is portrayed
in Fig.~\ref{fig:1}, and is described in detail in Sec.~\ref{sec:model}. Our
analysis focuses on the decay of the Majorana edge mode under the effect
of particle losses, starting from an initial state where such mode is occupied.
In Sec.~\ref{sec:homogeneous}, we first of all  study the case of a homogeneous
system as a benchmark, where we show that the Majorana mode decays exponentially
as a function of time. Specifically, we consider a particular setup where the
dissipation acts only on a part of the system (red area in Fig.~\ref{fig:1})
with a varying region near the boundaries that is not coupled to the Markovian
bath. This represents an idealized model system which, however, is potentially
relevant for a variety of experimental systems as we discuss in detail in
Sec.~\ref{sec:model}. Both in the presence and in the
absence of interactions, we find that for the case of particle losses the survival time of the Majorana edge
mode increases exponentially with the number of {\it protected sites}, that is,
the number of sites close to the edge where no dissipation occurs. This
increased lifetime can be immediately traced back to the nature of the edge
mode 
wave function which only exhibits an exponential tail extending into the bulk. 

In Sec.~\ref{sec:disorder} and Sec.~\ref{sec:lossgain}, we discuss how disorder
drastically affects the above mentioned
scenario. As anticipated before, in contrast to the exponential decay in the
homogeneous system, the Majorana edge mode decays as a {\it stretched
exponential} with a universal exponent $\alpha \simeq2/3$ for the pure loss dynamics. This points towards an increased robustness
of the edge mode when compared to the clean case. In particular, this universal
behavior occurs both in the Anderson and many-body localized regimes, i.e., also
in the presence of interactions which makes the system generic.
This shows that disorder, and, in general, localization effects, which induce
non-ergodic dynamics in the system, are not detrimental to the survival of
Majorana modes in the presence of a bath, and on the contrary, greatly enhance
their stability. 
This identifies disorder as a key mechanism to stabilize
topological features in open quantum systems, and thus to systematically improve
the stability of potential quantum 
memories even in the presence of dissipation which explicitly breaks the
symmetry protecting topological order. Finally, in Sec.~\ref{sec:lossgain}, we
present a case
study for the loss and gain case, where a qualitatively similar behavior can be
observed as well. We find that the decay of the Majorana again is of stretched exponential
form.

\section{Model Hamiltonian and master equation}
\label{sec:model}

The Kitaev chain can be realized in a variety of different experimental
architectures including AMO as well as solid-state setups
~\cite{alicea_majorana_solid_state,sau_majorana_solid_state,Beenakker:2011yg,
Goldstein:2011hl,Lutchyn:2010rw,Oreg:2010ud,mourik_majorana_nanowire,
Rokhinson:2012tw,Das:2012le, 
Brouwer25052012,Nadj-Perge31102014,jiang_majorana_driven_cold_atoms} 
\cite{Diehl:2011qp,Nascimbene:2012kq,PhysRevB.85.121405,PhysRevB.85.174533}. Its 
Hamiltonian describes
spinless fermions on a lattice of N sites with a pairing
term~\cite{kitaev_chain}: 
\be
H_\mathrm{K} = - \sum_{l=1}^{N-1} \left[ \left( J_l \ c_l^\dag c_{l+1} -\Delta_l
 c_l c_{l+1} \right) + \text{h.c.}  \right] + \mu  \sum_{l=1}^N c_l^\dag c_l,
\ee
where $c^\dagger_l,c_l$ are fermionic creation/annihilation operators at site
$l=1,\dots,N$. The first two terms represent tunneling between lattice sites as
well as pairing induced by a superconductor,
whose magnitude can, in principle, depend on the specific lattice site, and the
last term is a chemical
potential. In the homogeneous regime $J_l=J$, $\Delta_l=\Delta$,
$|\mu/J|<1,\Delta\neq 0$, the model
hosts a
symmetry-protected topological phase directly
related to the exact parity symmetry of the Hamiltonian, that is, $[H,P] = 0,P =
\prod_{l=1}^N (1-2n_l)$, where $n_l=c_l^\dag c_l$ is the local particle number operator. In
order to assure
that
the properties of the system are generic we account for a weak,
parity-preserving nearest-neighbor interaction making the system
non-integrable~\cite{Steinigeweg2013}. This yields as the full Hamiltonian:
\be
 H = H_\mathrm{K} + V,\quad 
 V = U \sum_{l=1}^{N-1} \left[n_l-\frac{1}{2}\right]
\left[n_{l+1}-\frac{1}{2}\right],
\label{eq:general_hamiltonian}
\ee
with $U/J \ll 1$. Notice that this model can be mapped via a Jordan-Wigner
transformation to a transverse-field XYZ chain, and as such is nonintegrable and
generic
\cite{Polkovnikov2011kx}. In the case where $\mu=U=0,
J=\Delta$, the ground state (GS) of the open Kitaev chain supports a localized
Majorana edge mode (see illustration in Fig.~\ref{fig:1}) whose wave function is
restricted to the edge lattice sites
only~\cite{kitaev_chain}. In the odd parity sector for odd number of sites, the
GS reads:~\footnote{We have checked that the qualitative behavior of the system is not affected by the initial state, unless the chain is initially empty.}
\begin{equation}
|\psi_0\rangle 
= 2^{-\frac{N}{2}}
\left[
\otimes_{i=1}^N \left( \ket{1}_i + \ket{0}_i \right) 
+  
\otimes_{i=1}^N \left( \ket{1}_i - \ket{0}_i \right) 
\right].
\label{eq:initial_condition}
\end{equation}
where the occupation of the zero energy mode, the edge-edge correlation,
$\theta$ satisfies:
\be
 \theta = \langle \psi_0 |\Theta |\psi_0 \rangle = 1,  \,\,\, 
 \Theta = i\left(c_1 + c_1^\dag \right) \left( c_N - c_N^\dag \right).
\ee
and $\Theta$ denotes the Majorana mode occupation operator. If we would slightly
perturb our Hamiltonian with $\mu,U > 0$, but still sufficiently small not to
leave the topological
phase, the Majorana zero mode develops an exponential tail extending into the
bulk, with most of its weight still located at the boundaries of the
chain\cite{brouwer_majorana_tail_bulk,stoudenmire_majorana_bulk}. 

The above particular choice of $|\psi_0\rangle$ represents an idealized
initial condition which, of course, relies on an exact implementation of the
Hamiltonian, or on a perfect dissipative state preparation of the state of
interest (along the lines of Ref.~\onlinecite{Diehl:2011qp}). As we would like
to argue at the end of this section, however, our full setup is still
sufficiently general in order to describe the generic dynamics.

Remarkably, it has been recently shown how the topological features of the
ground state in one-dimensional systems
- in particular, the presence of edge modes and a degenerate entanglement 
spectrum - can be extended to the entire spectrum in the presence of strong
disorder~\cite{Bahri2013,Chandran2014,Huse2013,bravyi2012disorder}.
This mechanism, a direct consequence of many-body localization, can lead to a
stabilization of
topological
features at nonzero energy density above the ground state in a closed
system.

Our main focus here is to investigate the fate of the Majorana zero mode
in the presence of symmetry- breaking dissipation. The microscopic details of
the particle loss and/or gain mechanism will strongly depend on the experimental
architecture. In a
cold atom experiment, losses are naturally occurring as either inelastic
scattering processes due to the immersion in a BEC of molecules, or background
collisions \cite{jaksch_cold_atom_losses,anglin_cold_atom_losses}.
These effects are often very well modeled as a Markovian
bath~\cite{Bloch:2008fc}.
In solid-state systems, loss and gain terms can emerge as a
consequence of a coupling to a grand-canonical bath. For Kitaev chains realized
by placing carbon nanotube wires on top of a p-wave superconductor, 
particle losses and gains can occur by electron tunneling into and from the
substrate. In this work, we concentrate on the
scenario where both the loss and gain can be described within a Master
equation framework. At a qualitative level, this implies that the parity
symmetry in the system
is violated {\it at all energy scales} - which we interpret as
a worst-case
scenario for symmetry-protected topological states.
The system dynamics can then be described using a Lindblad master equation of
the
form:
\begin{eqnarray}
\partial_t \rho &=& 
-\frac{i}{\hbar} \left[ H, \rho \right] 
+ \sum_{l=1}^{N} \kappa_l \ \left[ c_l \rho c_l^\dag - \frac{1}{2} \left\{
c_l^\dag c_l,\rho \right\} \right] + \nonumber\\
&+& \sum_{l=1}^{N} \gamma_l \ \left[ c^\dagger_l \rho c_l - \frac{1}{2} \left\{
c_l c_l^\dag,\rho \right\} \right],
\label{eq:master_equation_new}
\end{eqnarray}
with $\rho$ the density matrix of the system, $\kappa_l$ and $\gamma_l$ are the
local loss and gain 
rates, respectively. 
We consider uniform dissipation, but restricted to the bulk of the chain
where $\kappa_l=\kappa$, $\gamma_l=\gamma$ for $d <
l < N-d$, and $\kappa_l=\gamma_l=0$ otherwise with $d$ the distance of the first
dissipative site from the edge. For an illustration, see Fig.~\ref{fig:1}. In
other words, we mainly consider the influence of dissipation in the bulk, not at the boundaries.

The main
motivation for the choice of the setup is twofold. First, from a theoretical point of view, by restricting dissipation to the bulk we address the fundamental question of whether information stored in Majorana modes is stable against symmetry-breaking processes in the bulk. While dissipation acting on the
edges of the system opens a direct channel for the decay of the Majorana mode,
protecting the boundary sites allows us to extract the bulk
contributions and therefore to unravel the influence of the indirect channel. Second, our setup might be relevant also in a general experimental context where dissipation is not uniform and particle losses or gains are not occuring homogeneous throughout space. In that case, isolated regions with weak or even approximately vanishing decay rates can form for which our setup represents an idealized reduced model system.

We remark that we have checked via extensive numerical simulations whether the specific choice of 
the distribution of the decaying sites can  have an impact on the dynamics and our 
main results. We found that solely the number of sites which are not affected by dissipation
mattered in terms of the qualitative behavior - as long as the edge sites are not included.

The figure of merit in our study is the fate of the Majorana mode:
\begin{equation}
	\theta(t) = \mathrm{tr} [\rho(t) \Theta],
\label{eq:theta_t}
\end{equation}
as a function of time $t$ in presence of the incoherent bath.
For our initial density matrix $\rho(t=0) = | \psi_0\rangle \langle \psi_0|$ 
with $\theta=1$, we numerically solve the master equation in 
Eq.~\eqref{eq:master_equation_new}. For the noninteracting system with $U=0$,
the resulting master equation is of quadratic fermionic form and can therefore
be solved efficiently for large systems
\cite{Prosen2008,diehl_quadratic_form}. We use this
noninteracting limit as a benchmark for our more general and exact numerical
implementation where interactions can also be incorporated at the expense of
limited system sizes. We find, however, that, for the considered scenarios,
finite-size effects can be neglected (see discussion below).
We start by computing the dynamics of $\theta(t)$ in the clean case, and then
discuss the role of disorder, showing how the latter case displays to an
exponential gain in stability of the Majorana mode when compare to the former
one. 

Let us note that $\theta$ in Eq.~(\ref{eq:theta_t}) does not measure the
overlap with the idealized wave function of the Majorana mode with respect
to the final Hamiltonian for the general parameter set considered in this work
(the wave function will be generically localized at the edge, with a finite
contribution in the bulk), but rather those contributions which are located
right at the edges. This, however, is not a shortcoming as we will now shortly
discuss. First of all, the dynamics at the edges has to reflect the dynamics of
the full Majorana wave function because most of its weight is located right
there. Moreover, we expect that the considered scenario is also the
experimentally most relevant one. In particular, measuring operators located on
one lattice site is much easier than measuring an extended object such as the
full Majorana wave function. In this context, it is also important to emphasize
that the precise form of the Majorana wave function in a concrete experiment is
not a priori 
known because it depends on a variety of experimental details - especially
in the presence of interactions.

Finally, these arguments also motivate our specific choice of initial condition
in Eq.~(\ref{eq:initial_condition}). In any realistic experimental system it
will, of course, not be possible to precisely prepare the considered state due
to experimental imperfections. As long as the Majorana mode is initially
occupied, however, our observable $\Theta$ is capable to detect it without
knowledge about microscopic details. Therefore, we expect that these details
only lead to quantitative and not qualitative changes. 

\begin{figure}[t!]
\centering
\includegraphics[width=1\linewidth]{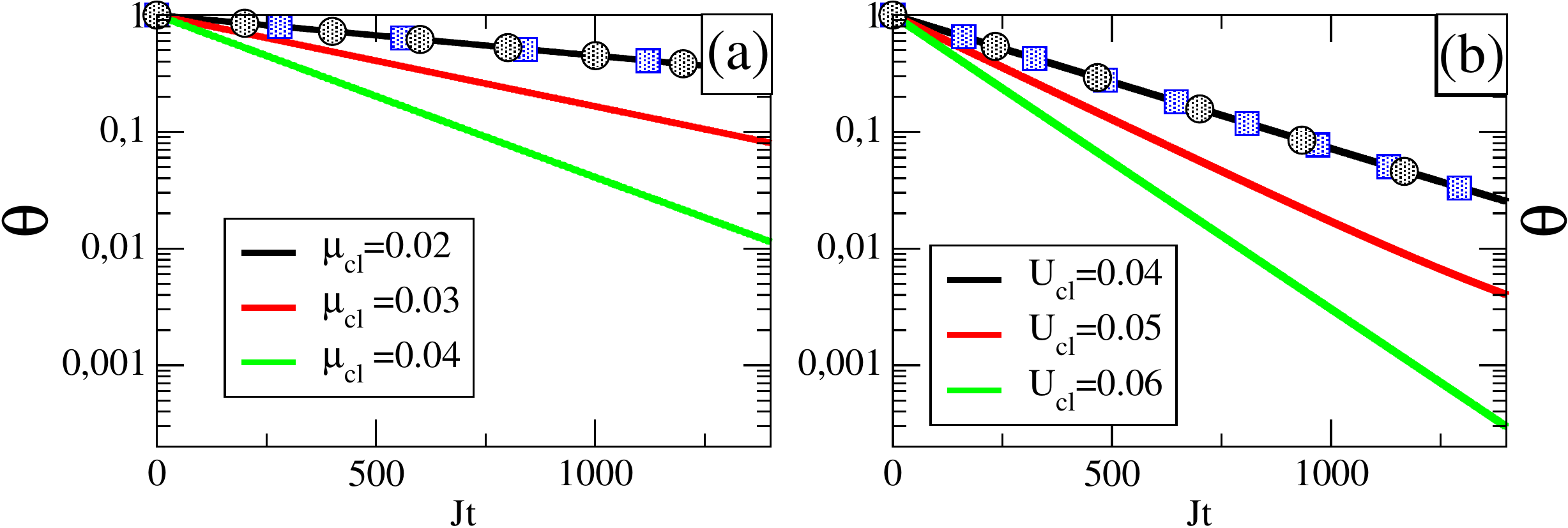}
\caption{Impact of particle loss on the edge-edge correlation $\theta$ in the
clean case $J_l=1$: (a) with a finite external field $\mu=0.02,0.03,0.04\
(U=0)$ and $\kappa=1/2$ for one protected site $d=1$; (b) with an on-site
repulsion $U=0.04,0.05,0.06\ (\mu=0)$ and $\kappa=1$ for $d=1$. Circles and
boxes in (a) and (b) show the dynamics for systems of different lengths
(N=9,11).}
\label{fig:results_diff_h_v_clean}
\end{figure}

\begin{figure}[b!]
\centering
\includegraphics[width=1\linewidth]{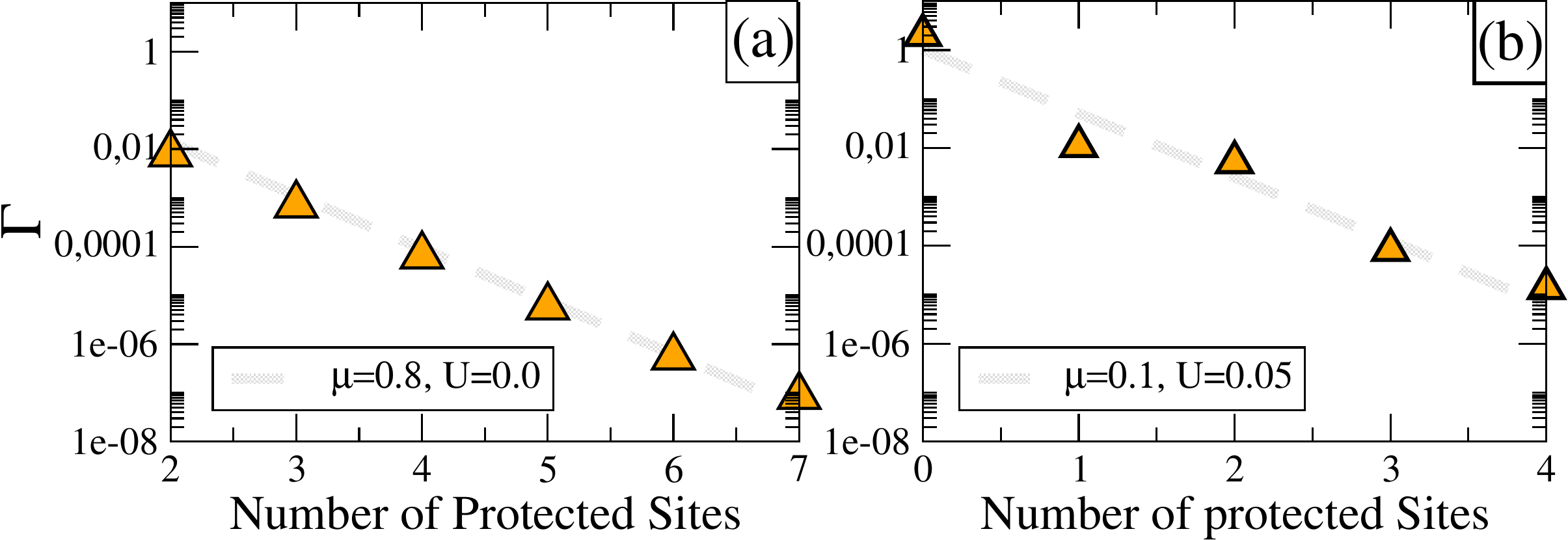}
\caption{Panels (c -d): exponential dependence of the
effective decay time $\Gamma$ on the number of protected sites $d$, with N = 43
(11) in the left (right) panel.}
\label{fig:protected_site_dependency}
\end{figure}
%


\section{Homogeneous system}
\label{sec:homogeneous}

Using the model system described in the previous section we now aim at studying
first the influence of symmetry-breaking particle losses for the homogeneous
system. This will serve as a benchmark for the further analysis in
Sec.~\ref{sec:disorder} where we will show that the Majorana edge mode can be
stabilized by the inclusion of disorder. For the homogeneous system we
analyze the noninteracting and interacting cases separately in order to single
out the potential influence of integrability-breaking interactions. 

The main result for the homogeneous system is that the Majorana mode occupation
decays exponentially in time in consequence of the breaking of the protecting
parity symmetry. However, we find that, remarkably, the associated decay rate
$\Gamma$ is highly sensitive to the distance $d$ of the first dissipative site
from the edge, but not to the microscopic parameters. In particular,
interactions do not lead to a qualitative change of the dynamics but rather only
increase the decay rates. Our main findings are summarized in
Fig.~\ref{fig:results_diff_h_v_clean}.

\subsection{Non-interacting case}
\label{sec:homogeneous_noninteracting}
We start by considering the benchmark of noninteracting particles with $U=0$. In
this limit we have used two different methods to solve the Master equation. On
the one hand we have made use of the property that the Master equation at $U=0$
becomes a quadratic form in terms of fermionic operators. This allows us to use
methods developed in Ref.~\onlinecite{Prosen2008}, such that we are able to
obtain the solution for large lattices up to $N=100$ sites. On the other hand we
have solved the Master equation directly on the basis of a Runge-Kutta algorithm
for up to $N=11$ lattice sites which can also be extended to interacting systems
studied later on. By comparing the results of both methods, we have found that
finite-size effects are negligible in the parameter regime considered here. 

As anticipated before, we find that in the non-interacting homogeneous system
the Majorana mode occupation $\theta(t)$ decays exponentially in time:
\be
    \theta(t) \stackrel{t\to\infty}{\longrightarrow} \,\, e^{-\Gamma t},
\ee
with $\Gamma$ the decay rate, both in the cases of a global bath acting on all
sites, $\kappa_l=\kappa$, and in the case of a bath which does not act on a
small region close to the edge, $\kappa_l=\kappa $ for $l\in [d+1, L-d]$ (with
$d$ the number of protected sites close to the edge), and 0 otherwise. In the
following, we focus on the bulk contribution, investigating the Majorana decay
as a function of $d$.

In Fig.~\ref{fig:results_diff_h_v_clean}a, time traces of the Majorana
occupations are shown for the case where the first dissipating lattice site is
located right next to the edge, i.e., $d=1$. As one can see, the decay rate
depends crucially on the strength of the chemical potential. This can be traced
back to a fundamental property of Kitaev chains at $\mu=0$. There the Majorana
qubit decouples completely from the bulk chain and Majorana mode occupation
operator $\Theta$ commutes with the Hamiltonian $[H_\mathrm{K},\Theta]=0$. As a
consequence, we have that $\theta(t)=1$ for all times $t$ when $\mu=0$.
Therefore, at this particular point in parameter space the Majorana completely
decouples from the dissipative dynamics and becomes completely stable. However,
this stabilization mechanism depends crucially on parameter fine-tuning and any
perturbation, always present in a realistic experiment, induces an exponential
decay.

The infinitesimal sensitivity of the Majorana qubit stability can be seen by
considering the effects of a non-vanishing chemical potential $\mu>0$, which is
shown in Fig.~\ref{fig:results_diff_h_v_clean}. Then, the Majorana qubit
experiences an effective coupling to the bulk modes and the exponential decay
induced by the symmetry-breaking losses sets in. The decay rate grows with
increasing chemical potential. 

Remarkably, it is possible to increase the lifetime of the Majorana mode
substantially also for the homogeneous system without parameter fine-tuning.
Specifically, we find that the decay rate is exponentially sensitive to the
distance $d$ of the edge from the first dissipative lattice site. This is shown
in Fig.~\ref{fig:results_diff_h_v_clean}, where $\Gamma$ is plotted as a
function of $d$. The decay rate is extracted numerically from two time points
$t_2 > t_1$ with $\Gamma =\text{Log}[\theta(t_2)/\theta(t_1)]/(t_1-t_2)$. We
have checked that the resulting $\Gamma$'s are independent of the precise choice
of $t_1$ and $t_2$ as long as we are in the asymptotic long-time regime. As a
consequence, in the case of pure particle losses, the bulk only contributes
weakly to the decay of the Majorana. Notice that this is different from the case
with losses and gain discussed in Sec.~\ref{sec:lossgain}.  We attribute
this strong sensitivity of the decay to the exponentially small overlap of the
Majorana wave 
function with the loss mechanism on distant lattice sites. In particular, as
long as the evolved Hamiltonian stays in the topological phase, the Majorana
wave functions  are exponentially localized in the vicinity of the edges.
Therefore, the direct influence of a particle loss at a given lattice site on
the Majorana mode is exponentially small. However, excitations that are created
by a particle loss could, in principle, still propagate to the edges potentially
inducing a decay of coherence. The numerical data, however, suggests that the
dominant principle is solely the overlap of the Majorana wave function on the
first dissipative lattice site.

\subsection{Interacting case}
\label{sec:homogeneous_interacting}

After having discussed in detail the noninteracting homogeneous system we now
aim at analyzing the influence of interactions. This is important for clarifying
which of the effects are generic and independent of particular parameter
choices. The numerical results for the interacting case are illustrated in
Fig.~\ref{fig:results_diff_h_v_clean}, where, along the lines of the previous
section, we check the resilience of the edge modes for increasing interaction
strength. 

As for the noninteracting case, we find that the Majorana edge modes decay
exponentially in time, see the numerical data presented in
Fig.~\ref{fig:results_diff_h_v_clean}. Again, the decay rate decreases
exponentially with the distance $d$ of the first dissipating lattice site, see Fig.~\ref{fig:protected_site_dependency}. In
contrast to the noninteracting case, however, an even-odd effect of $\Gamma$ as
a function of $d$ is visible. Importantly, the model is now generic which becomes apparent at the fine-tuned
parameter point $\mu=0$ of the noninteracting model, see
Sec.~\ref{sec:homogeneous_noninteracting}, where the Majorana mode is stable for
$d\geq 1$. In Fig.~\ref{fig:results_diff_h_v_clean}, we show time traces of
$\theta(t)$ at $\mu=0$ but nonvanishing interactions. As one can see,
interactions now lead to a decay of the edge mode. 

Before we study the impact of disorder, we summarize the main result of this
section:  Symmetry-breaking perturbations lead to a decay of the Majorana qubit
even if the edge states are protected from particle loss, independently on the
presence or absence of interactions. However, the survival time of the edge mode
increases exponentially with the number of protected sites $d$.

\section{Many-body localized Kitaev chain}
\label{sec:disorder}

For closed systems, it has been recently observed that disorder provides a
generic mechanism to stabilize topological order throughout the full many-body
spectrum, i.e., up to infinite temperature
\cite{li_topological_anderson_insulator_disorder_plateaus,
wu_robust_topological_insulator_disorder,Chandran2014}. 
Strikingly, this holds even for one-dimensional systems where long-range
(topological) order at nonzero temperature is not possible in thermodynamic
phases
\cite{rieder_majorana_disorder_phase_transitions,
basko_stabilization_disorder_interacting,
lobos_disorder_interaction_majorana_wires,vosk_mlb_1D}. 
These many-body localized systems are non-ergodic and consequently promise a key
mechanism to stabilize qubits in closed quantum system settings
\cite{franz_majorana_wires}.

For Anderson-localized systems it is known that a coupling to
low-temperature baths in general  induces a nonzero conductance in the context
of variable-range hopping~\cite{Mott1969}. However, depending on the spectral
details of the bath, localization can persist in the sense that the coupling
between system and bath becomes irrelevant at low energies~\cite{Hoyos2012}. For
small baths, not of thermodynamic nature, it has been shown that localization
can persist in a strong disorder limit~\cite{Huse2015,Nandkishore2015}. In many-body localized
systems the influence of low-temperature thermal baths onto the broadening of
local spectra has been studied recently~\cite{Nandkishore2014}. The static
properties of MBL systems coupled to small  baths and the transition of the
combined system to a Wigner-Dyson statistics have been addressed in
Ref.~\onlinecite{Johri2014}.

\begin{figure}
\centering
\includegraphics[width=\linewidth]{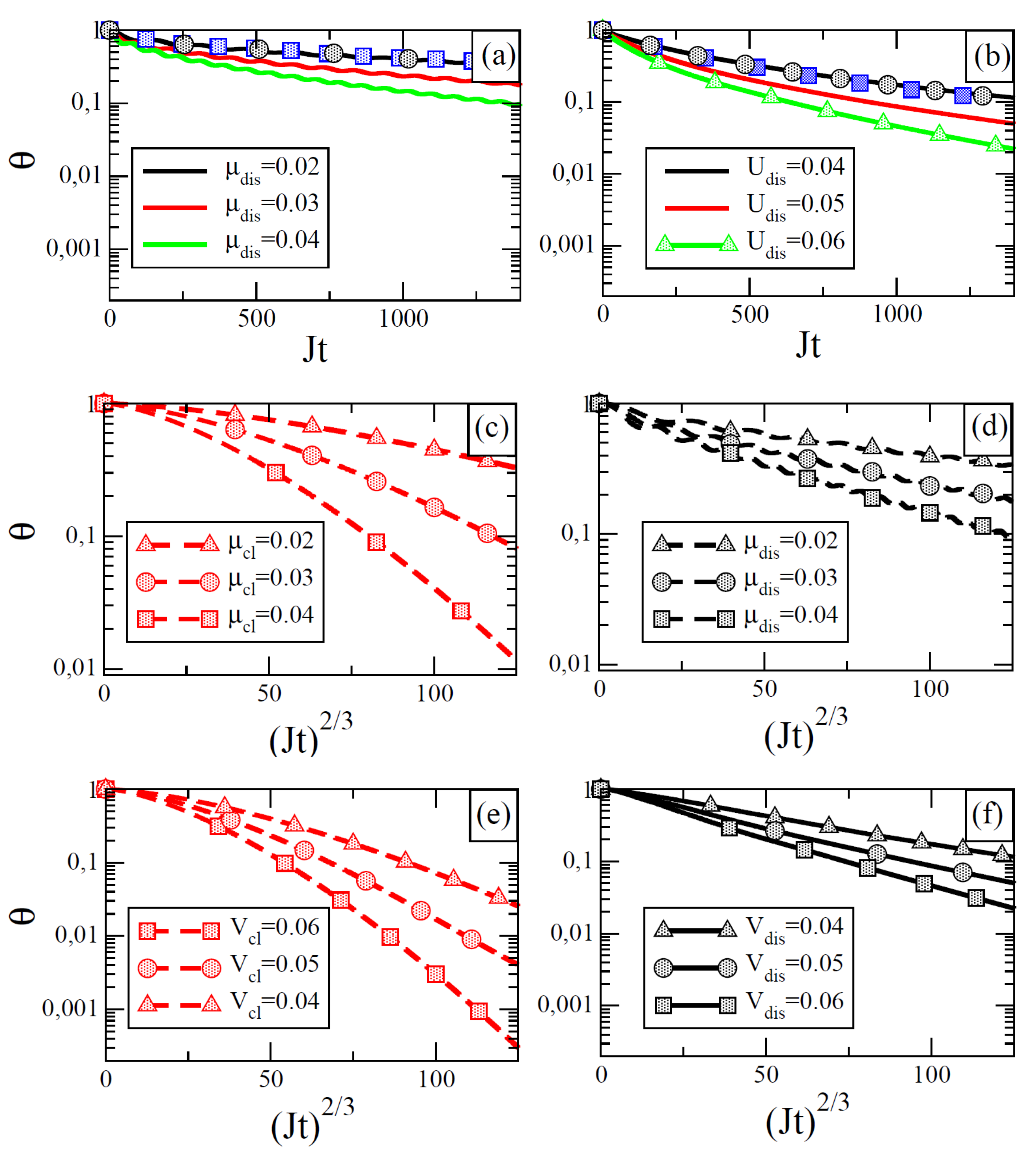}
\caption{Impact of particle loss on the edge-edge correlation $\theta$ in the
disordered case $J_l\in[0,e]$: (a) with a finite external field
$\mu=0.02,0.03,0.04\ (U=0)$ and $\kappa=1/2$ for one protected site $d=1$; (b)
with an on-site repulsion $U=0.04,0.05,0.06\ (\mu=0)$ and $\kappa=1$ for $d=1$.
Circles and boxes in (a) and (b) show the 
dynamics for systems of different lengths (N=9,11). Panels (c-f) show the time
trace with a scaled time-axis to illustrate $t^{2/3}$ dependence. In panels (c)
and (e), we show the clean case, where the decay is exponential, and thus, in
the rescaled axis, an algebraic decay with power larger than 1 is observed. In
panels (d) and (f), we contrast this with the disordered case, where the decay
is a stretched exponential with a power $\alpha\simeq 2/3$.}
\label{fig:results_diff_h_v_disordered}
\end{figure}

In the following, we study the many-body
localized Kitaev model in presence of an incoherent structureless bath described
by the Master equation in Eq.~(\ref{eq:master_equation_new}). We show the
stabilization property of many-body localized phases remarkably extends also to
this worst-case open quantum system scenario, thus
identifying non-ergodicity of closed systems as a possible route to stabilize
edge states in the presence of detrimental (symmetry-breaking) noise. 
Our main result,
summarized in Fig.~\ref{fig:results_diff_h_v_disordered}, is that the Majorana
mode can be stabilized by the inclusion of disorder with an exponential gain in
resilience. Specifically, the exponential decay of the homogeneous system
converts into a stretched exponential decay in presence of disorder, that is: 
\be
\langle \theta(t) \rangle_\mathrm{dis} \stackrel{t\to\infty}{\longrightarrow}
e^{-(\gamma t)^\alpha}
\label{eq:stretched_exponential_decay}
\ee
with $\langle \dots \rangle_\mathrm{dis}$ denoting the average over disorder
realizations. We find that within our numerical accuracy the associated exponent
$\alpha$ is universal and close to
\begin{equation}
	\alpha \simeq 2/3.
\end{equation}
In our numerical implementation, we choose bond disorder for simplicity by
introducing inhomogeneous couplings $J_l \in [0,W]$, $l=1\dots N-1$ sampled from
independent uniform distributions of width $W =eJ$. We will characterize the
strength of the disorder via $J$ differing from $W$ by Euler's number $e$ such
that the critical point of the noninteracting model is located at $\mu/J=1$ as
in the clean case \cite{Pfeuty1979,Fisher1995}. In the following simulations, we are interested in the large disorder limit such that the localization length of most eigenstates is limited to few lattice spacings (thus limiting finite-size effects). For each parameter
regime, the results are averaged over up to $10\,000$ disorder
realizations in order to ensure convergence. We expect, however, that the
generic features are independent of this precise choice of bond disorder, in
particular, because it is known that the properties of the bond-disordered
Kitaev and the related Ising chain are equivalent to the case of site disorder.

Before diving into a detailed discussion of our results, let us comment on the
case where the dissipative losses act on every lattice site, $\kappa_l=\kappa$,
in 
particular, including the edges where the Majorana modes are located. In this
case, we find that the Majorana mode occupation still decays exponentially as
found for the homogeneous case. Thus, the Majorana qubit is not stable against
incoherent onsite losses as one might expect. This local decoherence mechanism
cannot be stabilized using disorder. However, as we will show below, protecting
sites in the immediate vicinity of the boundaries lead to a substantial
stabilization.

We start by discussing as a warm-up the noninteracting (Anderson) limit with
nonzero chemical potential $\mu>0$ but $U=0$. Our numerical results on this case
are shown in Fig.~\ref{fig:results_diff_h_v_disordered}a. In panels
\ref{fig:results_diff_h_v_disordered}c-d, we plot the same quantity for a clean
(c) and disordered system (d, same data as in a) as a function of the rescaled
time $(Jt)^{2/3}$. While in the clean system the decay stays exponential as
discussed in the previous section (the curve bends down in lin-log scale), in
the disordered case one gets an almost flat line (but for additional coherent
oscillations), demonstrating an exponentially increased stability of the edge
mode, and a decay according to $\alpha\simeq 2/3$. The decay becomes faster for
larger chemical potential as this increases the coupling between the Majorana
qubit and the bulk modes. An analysis of the time derivatives of the Majorana
mode survival also confirms this decay.

The same stabilization mechanism emerges in the many-body localized regime, where the 
Majorana operator is expected to be renormalized, still be localized at the boundary~\cite{Ros:2014kq}. The
resulting decay of the Majorana mode for different interaction strength for the
case $\mu=0$ is illustrated in Fig.~\ref{fig:results_diff_h_v_disordered}b, f,
while panels (e-d) show the comparison between clean and disordered case. Again,
the decay is of the form of Eq.~(\ref{eq:stretched_exponential_decay}) with
$\alpha \approx 2/3$.

\begin{figure}[t]
\centering
\includegraphics[width=\linewidth]{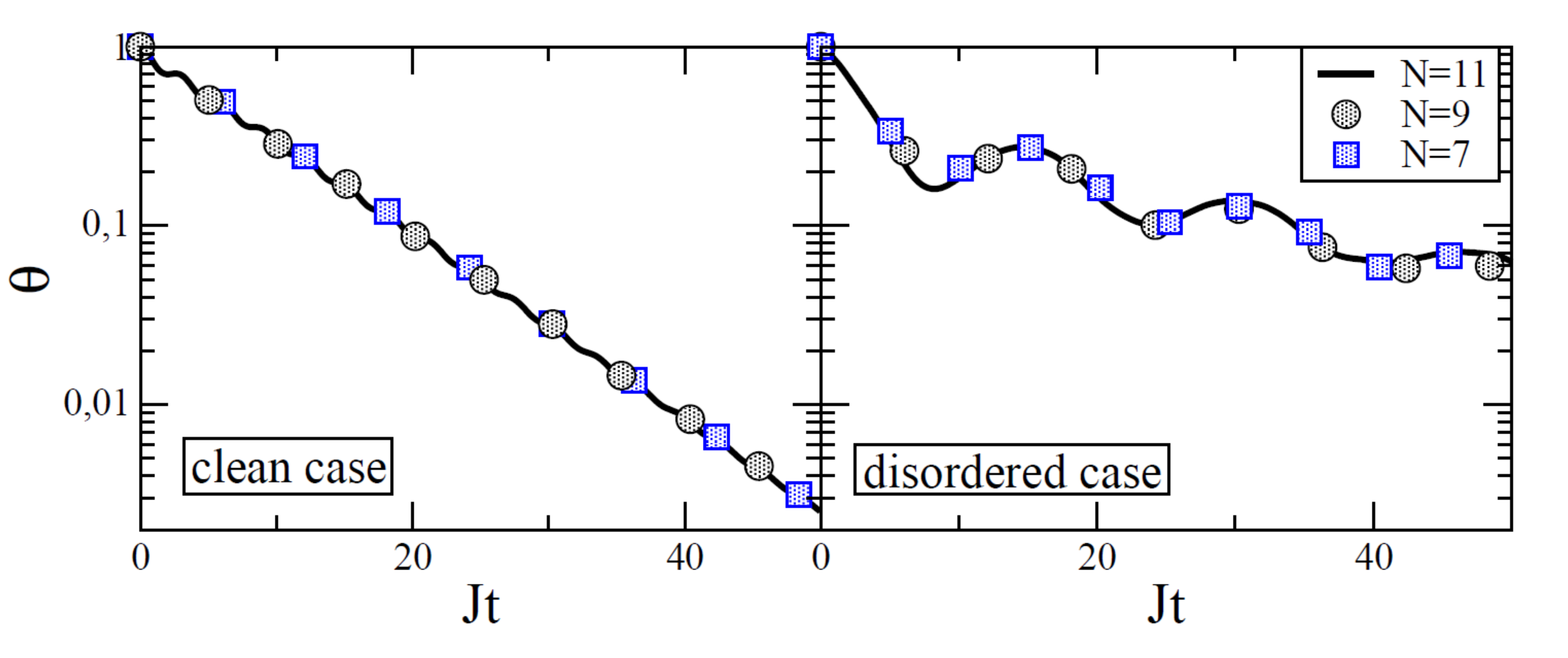}
\caption{Limit of strong interaction and external field $\mu=U=0.2$ with d=2
protected sites and $\kappa=1.5$ for the clean and disordered case. Here,
disorder increases the oscillation amplitude and slows down in order of
magnitude the decay of the edge-edge correlation.}
\label{fig:results_enhanced_coherence}
\end{figure}

Finally, let us consider the case of both nonzero interactions and nonzero
chemical potentials. The resulting temporal behavior of the Majorana mode is
shown in Fig.~\ref{fig:results_enhanced_coherence}. Here we consider a
comparable strength for both $U=J/5$ and $\mu=J/5$ at a dissipation strength
$\kappa/J=3/2$. For comparison also the respective homogeneous case is shown
that exhibits the pure exponential decay already discussed in
Sec.~\ref{sec:homogeneous}. The many-body localized case on the other hand again
shows the stretched exponential decay, with an additional coherent oscillation
on top of the decay is found. 

\section{Effects of particle gain}\label{sec:lossgain}

So far, we concentrated on a setup where only  particle losses occur. 
Such as scenario corresponds mainly to potential  implementations based on cold
atom setups, where the main
dissipation channel is given by inelastic scattering between molecules in the
reservoir, and particles in the wire. 
In a condensed-matter setting, a more natural environment would generically be a
grand-canonical bath inducing both particle losses and gains
\cite{su2013collective,callsen2013steering}. This is the scenario we focus on in
this
section. Specifically, we consider equal rates
$\kappa_l=\gamma_l$ for both losses and gains for simplicity. Of course, baths
in typical condensed-matter systems will not be exactly of the type described by
the Master equation in Eq.~(\ref{eq:master_equation_new}) except maybe
high-temperature baths. As anticipated in Sec.~\ref{sec:model}, however, from
the perspective of the stability of the Majorana mode, one can interpret the
dynamics induced by this Master equation with {\it incoherent} losses and gains
as a worst-case scenario.

\begin{figure}
\centering
\includegraphics[width=\linewidth]{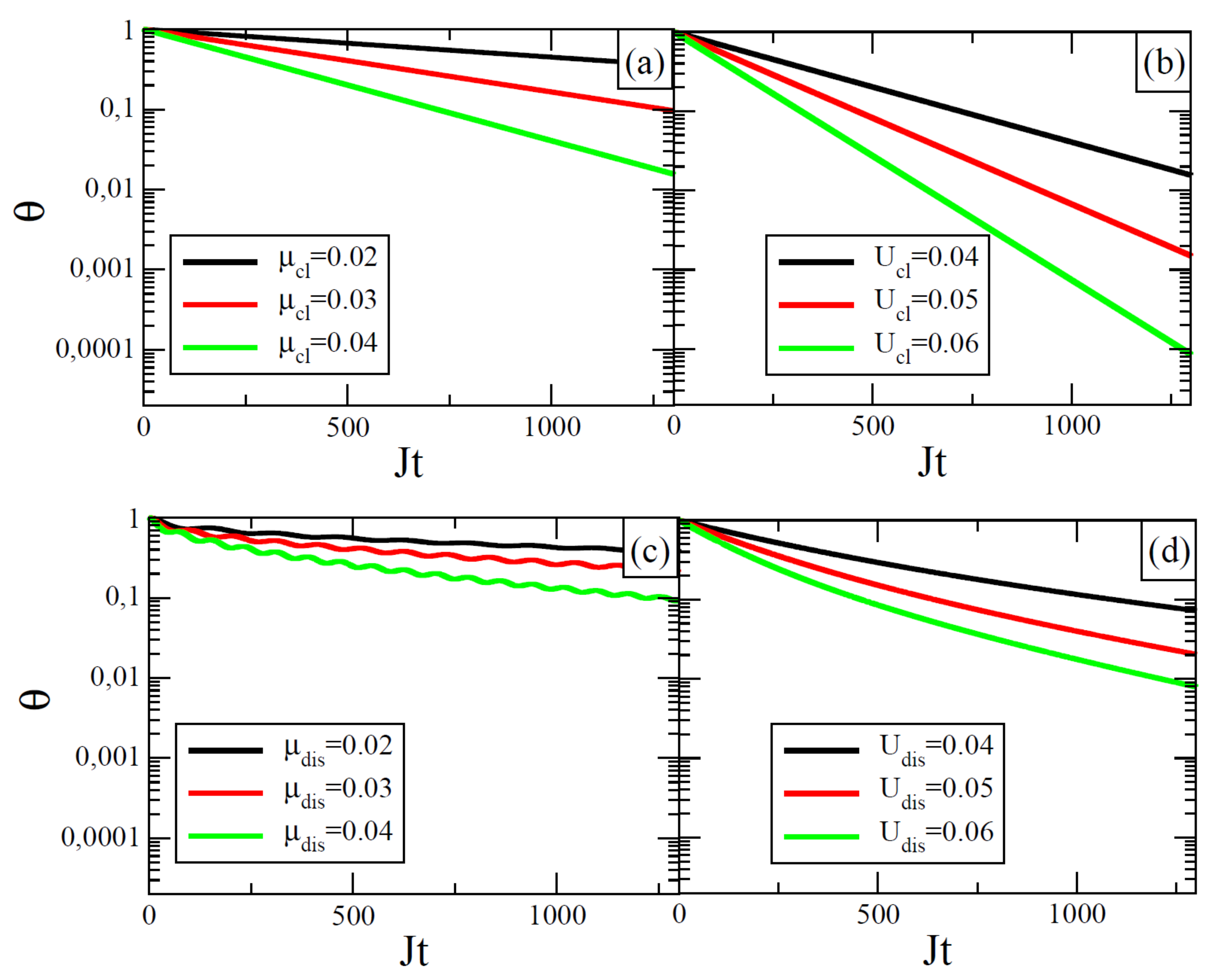}
\caption{Impact of particle loss and gain on the edge-edge correlation $\theta$
in the clean (upper panel: a,b) and disordered case (lower panel: c,d)
$J_l\in[0,e]$: (a) with a finite external field
$\mu=0.02,0.03,0.04\ (U=0)$ and $\kappa=1/2$ for one protected site $d=1$; (b)
with an on-site repulsion $U=0.04,0.05,0.06\ (\mu=0)$ and $\kappa=1$ for $d=1$.}
\label{fig:results_gain_and_loss_case}
\end{figure}

We proceed as before by comparing the homogeneous case with the disordered case
for
a non-interacting, i.e. $U=0$, and interacting chain with $U\neq0$. We will
focus on a
single protected site at each edge for simplicity.
Our main results for this setup are summarized in 
Fig.~\ref{fig:results_gain_and_loss_case}.
In principle, the overall picture remains identical  compared to the case with
pure losses.
Again, we find that the presence of disorder suppresses the decoherence process
of the Majorana mode. At first glance, also the qualitative picture is not
changed, as we find a
clear exponential decay in the clean case and a stretched exponential in the
disordered case. This is not surprising, as from a symmetry point of view, both
particle
loss alone and particle loss and gain break the parity symmetry and in 
consequence lead to a loss of coherence of the Majorana edge mode.

However, apart from the first observation, that the additional particle
gain does not completely change the overall impact of disorder on the
decay of the Majorana, we find subtle modifications in the universality of the
stretched exponential scaling. In contrast to previous case with pure losses
where the exponent has been $\alpha \simeq 2/3$, see
Eq.~(\ref{eq:stretched_exponential_decay}), we find a modified exponent $\alpha$
now in the presence of the additional particle gains. Specifically, from our
numerical simulations the exponent turns out to be close to $\alpha \approx
3/4$. However, we have to emphasize that the situation is much less clear
compared to the pure loss case. In particular, we find some slight dependence on
the microscopic parameters. Therefore, we cannot conclude from our data that
the exponent is universal in this case. \footnote{Whether the decaying exponent can
become universal for larger systems cannot be addressed with the 
methods employed here.}
For many cases, we find a stretched exponential decay with
approximately $\alpha=3/4$, but this value can vary (for the parameters
investigated) within a range $\pm 10\%$.

Furthermore, it is worth mentioning that the balanced particle loss and
gain leads to a more complex picture, compared to the pure loss scenario. For
example, we observed partially decoupled bulk and edge dynamics in the case of
an interacting chain of length $N=7$, where the distance exponential decay in
the homogeneous case does not depend on the number of protected sites. This
feature is a result of the uniform bulk dynamics. As soon as the bulk settles to a steady state due to gain and loss, the absolute length of the bulk is not of
importance anymore. This stands in contrast to the inhomogeneous case, where
such a distance dependence is tractable.

\section{Conclusions and outlook}
We have presented a numerical study of the influence of a symmetry-breaking
Markovian bath on the stability of Majorana edge modes in a Kitaev chain,
including both disorder and interactions. In a clean setting, the occupation of
the Majorana mode decays always exponentially as a function of time regardless
of the system being interacting or not. The corresponding decay rate $\Gamma$
displays an exponential dependence on the number of sites close to the edge
where dissipation is not acting, a direct consequence of the exponential
localization of the mode wave-function in the pure loss case.  This indicates that protecting even a very small number of sites can
already improve the Majorana mode lifetime by several orders of magnitude with
respect to the non-protected case, if gain is negligible or other
interactions prevent an effectively reduced bulk length.

Strikingly, an even more solid protection mechanism is provided by the presence
of disorder. In the closed system case without dissipation, many-body
localization in disordered systems prevents thermalization even in the presence
of interactions~\cite{Nandkishore2014,Altman2014}. Importantly, it has been
shown that topological properties can be present over the entire Hamiltonian
spectrum in many-body localized systems~\cite{Bahri2013,Chandran2014}, and 
that bulk transport is severely inhibited in such phases - which generically display vanishing 
conductivity.
We found that the consequences of these disorder-induced mechanisms upon the introduction of a Markovian bath are that the Majorana mode occupation decays in a qualitative
different way than in the clean case. In the case of loss-only dissipation,
which is directly relevant to cold atom systems, the decay form is a stretched
exponential, with universal power $\simeq 2/3$, while in the case of both loss
and gain - closer to solid state scenarios - the decay is still compatible with
a stretched exponential, but with a larger power close to $\simeq 3/4$, whose 
determination is generically more challenging. 

Overall, our results point toward the possibility of using many-body
localization as a stabilizer mechanism of edge states in symmetry-protected
topological states in the presence of a bath with explicitly breaks the
symmetry. While the results presented herein are obtained using a controlled
numerical approach, it would be of immediate relevance to develop a more general
understanding using approximate solutions, which capture the main features of
many-body localized systems coupled to baths, e.g., by exploiting the emergence
of approximate integrals of motion, or by treating the bulk as an effective bath for the edges. 
Moreover, while the work here deals with
symmetry-protected topological phases, the addition of disorder could serve as a
stabilizing mechanism under dissipation even in phases supporting topological
order in combination to long-range 
entanglement~\cite{wenbook,PhysRevB.83.075102,PhysRevB.81.134509}, where it has
been recently shown how disorder can be used to localize {\it unwanted}
excitations~\cite{Stark2011,Wootton2011}, and could also help in further stabilizing 
Majorana modes against symmetry-preserving noise\cite{Pedrocchi:2015eu,Hu:2014ly,Goldstein:2011hl,Hu:2015dz}. 

\acknowledgements 
 We acknowledge useful discussions with M. Baranov,
E. Demler, M. Lukin, A. Scardicchio, and N. Yao, and we thank P. Zoller for fruitful criticism.
Work in Innsbruck is partially supported by the ERC Synergy Grant UQUAM, SIQS,
and SFB FoQuS (FWF Project No.~F4016-N23). M. H. has been supported by the
Deutsche Akademie der Naturforscher Leopoldina via Grant No. LPDS 2013-07 and LPDR 2015-01.
A. C. acknowledges gratefully support from Alexander-von-Humboldt foundation
through the Feodor-Lynen program.

\bibliography{references}
\end{document}